\documentclass[runningheads]{llncs}
\usepackage[english]{babel}
\usepackage{cite}
\usepackage[letterpaper,top=2cm,bottom=2cm,left=3cm,right=3cm,marginparwidth=1.75cm]{geometry}
\usepackage{amsmath}
\usepackage{fancyhdr}
\usepackage{graphicx}
\usepackage{booktabs}
\usepackage{float}  % 加入导言区
\usepackage[colorlinks=true, allcolors=blue]{hyperref}
\usepackage[font=small]{caption}
\usepackage[font=footnotesize]{caption}
\usepackage{graphicx}
\usepackage{subfigure}
\title{Heterogeneous network drug-target interaction prediction model based on graph wavelet transform and multi-level contrastive learning}
\author{Wenfeng Dai\and Yanhong Wang\and Shuai Yan\and  Qingzhi Yu\and Xiang Cheng\thanks{Correspondence: School of Information Engineering, Jingdezhen Ceramics University,mail:jx\_chx@126.com} }
\institute{Jingdezhen Ceramics University,Jiangxi,China}
\begin{document}
\pagestyle{fancy}
\fancyhead{}
\maketitle

\begin{abstract}

Drug-target interaction (DTI) prediction is a core task in drug development and precision medicine in the biomedical field. However, traditional machine learning methods generally have the "black box" problem, which makes it difficult to reveal the deep correlation between the model decision mechanism and the interaction pattern between biological molecules. This study proposes a heterogeneous network drug-target interaction prediction framework, integrating graph neural network and multi-scale signal processing technology to construct a model with both efficient prediction and multi-level interpretability. Its technical breakthroughs are mainly reflected in the following three dimensions: (1) Local-global feature collaborative perception module. Based on heterogeneous graph convolutional neural network (HGCN), a multi-order neighbor aggregation strategy is designed. (2) Multi-scale graph signal decomposition and biological interpretation module. A deep hierarchical node feature transform (GWT) architecture is proposed. (3) Contrastive learning combining multi-dimensional perspectives and hierarchical representations. By comparing the learning models, the node representations from the two perspectives of HGCN and GWT are aligned and fused, so that the model can integrate multi-dimensional information and improve the prediction robustness. Experimental results show that our framework shows excellent prediction performance on all datasets. This study provides a complete solution for drug target discovery from "black box prediction" to "mechanism decoding", and its methodology has important reference value for modeling complex biomolecular interaction systems. The source code is available at \url{https://github.com/chromaprim/SHGCL-DTI-GWT}.

\end{abstract}
\section{Introduction}
Drugs, as small molecule preparations that regulate the function of target proteins, are the core means of treating diseases in modern medicine. However, the new drug development process from target identification to clinical launch takes 10-15 years and an average investment of US \$2.6 billion. Drug-target interaction (DTI) prediction is a key link that directly affects the efficiency of lead compound screening. Taking anti-tumor drug development as an example, only 3\% of preclinical candidate compounds are ultimately approved, and improving the accuracy of DTI prediction can directly reduce R\&D costs by 40\%\cite{martin2017much}. Computer-aided drug design (CADD) has significantly changed the traditional trial-and-error R\&D model by constructing a "structure-activity" mapping model:Experimental data driven: Integrate multi-dimensional biomarkers (including compound physicochemical properties, protein binding affinity, metabolic stability spectrum, etc.) to construct feature space;

Computational modeling breakthrough: The DTI prediction field has formed three main research directions:
(1)Structure-based computational methods: Molecular docking and free energy calculations are used to simulate the geometric conformation of drug-target interactions, but the computational cost is high \cite{xiong2023improving,asada2021using,ar2024deep}. Baohua Zhang\cite{zhang2022molecular} used structure-based virtual screening methods to identify potential highly active compounds, thereby accelerating the process of new drug design. Qun Su et al.\cite{su2025robust}calculated the absolute binding free energy by integrating physical laws and geometric knowledge to establish a robust protein-ligand interaction model.
(2)Phenotype-based deep learning methods: Graph neural networks (GNNs) and Transformer architectures are used to directly learn nonlinear rules from compound activity data \cite{mitchell2023proteome,li2023drugmap,kayikci2018visualization}. Torng W et al. \cite{torng2019graph}trained two Graph-CNN models to automatically extract features from pocket graphs and 2D ligand graphs for protein-ligand interaction prediction. Kexin Huang et al. \cite{huang10molecular} used an enhanced Transformer encoder to extract semantic relationships from a large amount of unlabeled biomedical data substructures. Abbasi, M. et al. used generative adversarial networks to design optimized drug candidates\cite{abbasi2022designing}.
(3)Hybrid methods: integrating quantitative structure-activity relationship (QSAR) with deep learning, such as the SwissADME prediction model that simultaneously optimizes ADMET properties such as absorption and distribution \cite{tropsha2024integrating,er2023design,er2023qsar,bakchi2022overview}. Hongjie Wu et al.\cite{wu2024attentionmgt}proposed a multimodal attention-based DTA prediction model, AttentionMGT-DTA, which uses molecular graphs and binding pocket graphs to represent drugs and targets, respectively, and adopts two attention mechanisms to integrate and interact information between different protein modalities and drug-target pairs.

Despite the significant progress of CADD technology, deep learning-driven DTI prediction still faces three challenges: data bias problem: the ratio of positive and negative samples in the DTI dataset is seriously unbalanced (usually <1:100), which leads to overfitting of the model in unseen compounds (ROC curve deviates significantly in the low activity region); interpretability dilemma: the black box model cannot quantify the contribution of key residues to binding energy; lack of dynamic characteristics: existing models mostly use static protein structures and fail to capture the impact of dynamic changes in target conformation on binding strength.

This paper proposes a DTI prediction framework that integrates multi-scale features of heterogeneous graphs to solve existing bottlenecks through the following innovations:
(1)Multi-scale feature extraction: Design a graph wavelet transform module to capture the evolutionary conservation characteristics of protein domains through the low-frequency filter of graph wavelet transform, and design a high-throughput filter to identify the characteristics of dynamic conformational changes.
(2)Heterogeneous data fusion: Construct heterogeneous graphs to integrate compound molecular graphs (nodes: atoms; edges: chemical bonds), protein structure graphs (nodes: residues; edges: spatial distances) and bioactivity data, and achieve multimodal feature alignment through cross-graph attention mechanisms. Dynamic context perception and multimodal alignment are achieved through semantic attention mechanisms, which significantly improves the performance and interpretability of tasks such as drug discovery and biological text analysis.
(3)A cross-view multi-level contrastive learning framework is constructed to achieve multi-dimensional feature alignment through a three-stage contrastive learning mechanism:View generation: generating node representations from HGCN (topological view) and GWT (frequency domain view) respectively;Feature matching: using InfoNCE Loss to maximize the similarity of features of the same node under different views; Knowledge distillation: designing an intermediate teacher network based on molecular docking energy to map complex features into an interpretable biophysical space.

\section{Dataset}
To verify the prediction effect of the model, we selected Yunan Luo's dataset, which contains four types of nodes and eight different types of edges. 
The nodes are proteins, drugs, diseases, and side effects, with the number of nodes 
\begin{figure}[htbp]
    \centering
    \includegraphics[width=0.9\textwidth]{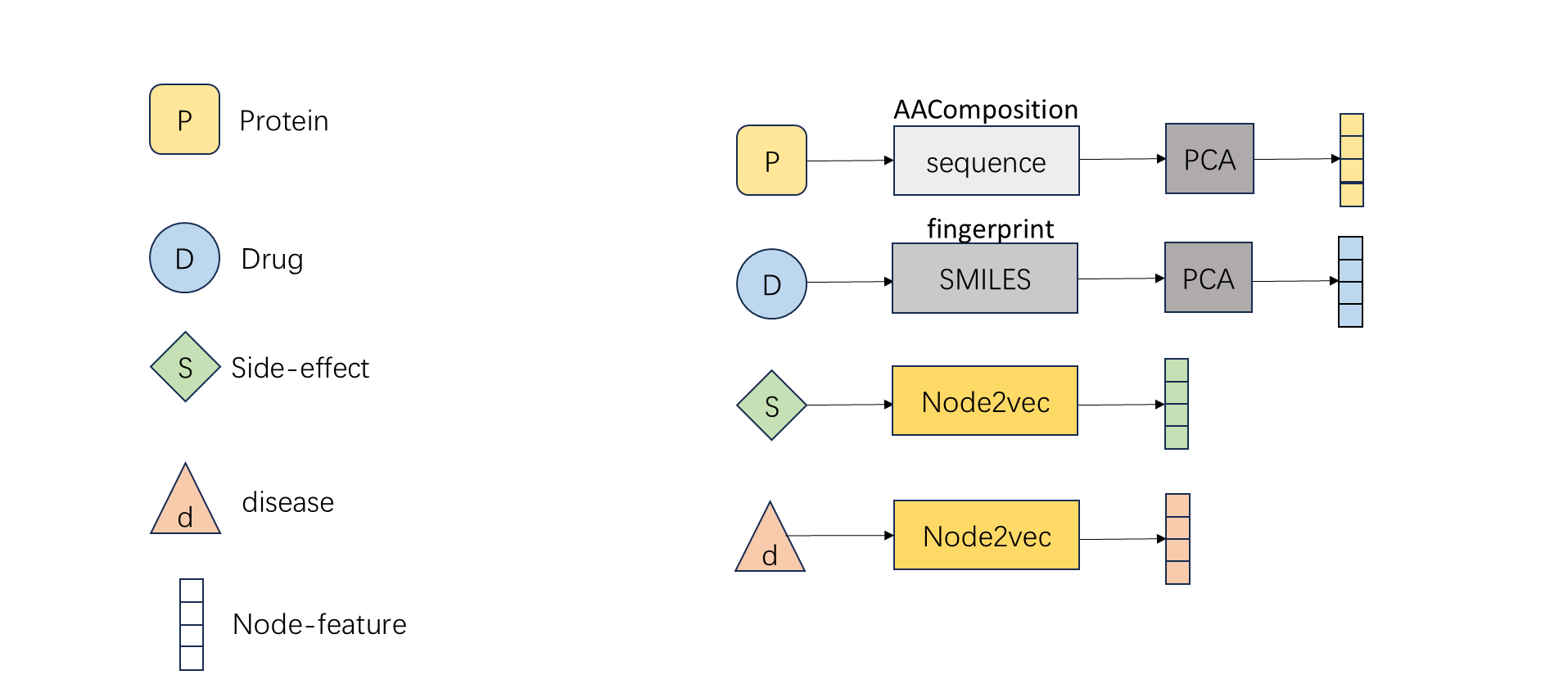}
    \caption{\label{fig:main1} Feature acquisition of four types of nodes using PCA and Node2vec}
    \label{fig:main1}
\end{figure}
being 1512, 708, 5603, and 4192 respectively. The edges are drug-protein interactions, drug-drug 
interactions, drug-disease interactions, protein-protein interactions, protein-disease interactions, drug-side effect interactions, drug-drug similarity, and protein-protein similarity\cite{luo2017network}.
The node feature extraction is shown in Figure \ref{fig:main1}. For drug nodes, the SMILES molecular structure is extracted and converted into a molecular fingerprint, and then the principal component analysis \cite{dunteman1989principal} is used to reduce the dimension to obtain a low-dimensional vector representation of the drug.The protein node is based on the amino acid sequence, and its amino acid composition characteristics are calculated. The feature representation is also obtained after PCA dimensionality reduction. The side effect and disease nodes are represented by the Node2vec method \cite{grover2016node2vec}.

In order to integrate different types of nodes and edges to form a complex heterogeneous network structure, a self-loop edge is created for each node type based on the above eight edges to represent the relationship between the node and itself.Drug-drug similarity and protein sequence data are processed by threshold filtering to remove low-weight edges and ensure that only significant relationships are retained. Finally, a heterogeneous network for drug-target interaction prediction is constructed using the above nodes and edges.

\section{Model Architecture}
In order to effectively predict the relationship between protein-target interactions, we designed a multi-perspective heterogeneous network graph convolution model: first, graph convolution learning is performed from the neighbor perspective and the deep perspective on the constructed heterogeneous network. The former focuses on capturing the direct neighbor structure of the node, and the latter uses cross-type multi-hop paths to mine the deeper node relationship. Then, the node representations from the two perspectives are aligned and fused through multi-level contrastive learning; finally, the drug-protein interaction matrix is predicted based on the fused features to achieve accurate prediction of protein-target interactions. The overall architecture of the model is shown in Figure \ref{fig:main2}, which mainly consists of four parts: neighbor perspective encoding HGCN \cite{zhu2020hgcn} (SC Encoder), deep perspective encoding GWT (MG Encoder), multi-level contrastive learning and drug target prediction.
\begin{figure}[hbtp]
    \centering
    \includegraphics[width=0.9\textwidth]{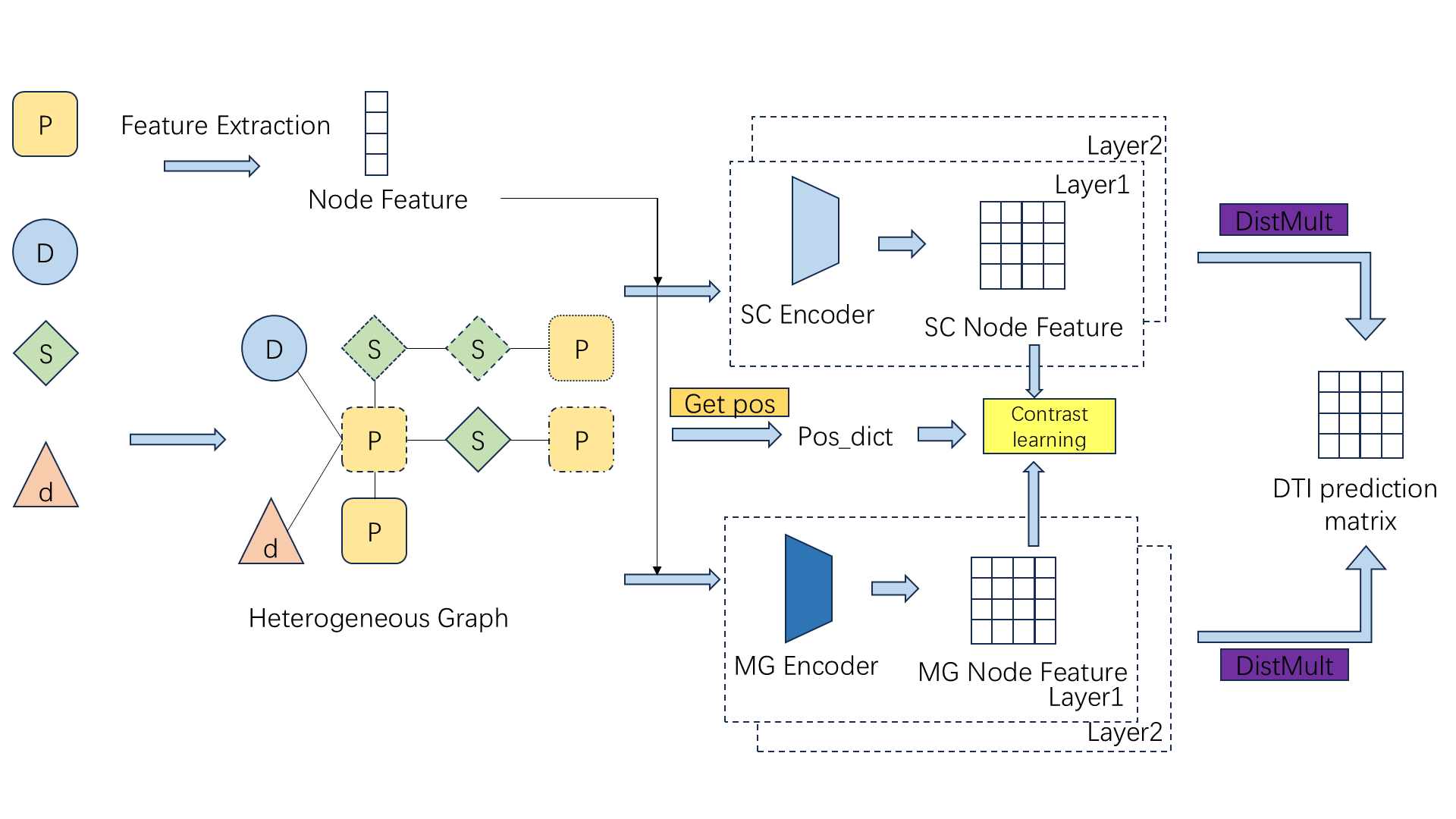}
    \caption{\label{fig:main2} GHCDTI overall model architecture}
    \label{fig:main2}
\end{figure}

\subsection{Neighborhood Node View Encoding (SC Encoder)}
In order to extract the characteristic information of drug-target interaction from the constructed heterogeneous network, we use HGCN to aggregate neighbor node information. HGCN is a neural network model specially designed to process graph structured data. It learns the characteristic representation of nodes by propagating information between neighbor nodes in the graph, thereby capturing the structural information of the graph and the relationship between nodes. In heterogeneous networks, node features are updated by aggregating information from neighbors of different types. We use two layers of HGCN to propagate and aggregate the features of each node type.The formula for aggregating node information through HGCN is as follows:
\begin{equation}
H_v^{i} = \frac{1}{|N(v)| + 1} \left( \sum_{u \in N(v)} \widetilde{D}_{v,u}^{-\frac{1}{2}} \widetilde{A}_{v,u} \widetilde{D}_{v,u}^{-\frac{1}{2}} H_u^{i} W_{v,u} + H_v \right)
\end{equation}

Among them, $N(v)$ represents the neighbor node type set of node type $v$. $A_{v,u}$ is the relationship between node types $v$ and $u$. $\widetilde{A}_{v,u} = A_{v,u} + I$ indicates the added adjacency matrix. $\widetilde{D}_{v,u}$ is the degree matrix of $\widetilde{A}_{v,u}$, used for normalization. $H_u$ is the feature representation of neighbor node type $u$, and $W_{v,u}$ is a learnable weight matrix used to perform linear transformation on neighbor node features. $H_v$ is the inherent feature of node type $v$. $\frac{1}{|N(v)|+1}$ is the normalization coefficient.

To ensure that the aggregated features remain numerically stable, we stack two encoding layers so that nodes can aggregate the information of their two-hop neighbors. For node, it can capture more extensive graph structure information.

\subsection{Deep View Encoding (MG Encoder)}
In order to mine deep associations in cross-type multi-hop paths from the constructed heterogeneous network, we adopt the Graph Wavelet Transform (GWT) module as the heterogeneous network perspective encoder. Different from HGCN that directly aggregates neighbor information, GWT processes graph signals through multi-scale feature decomposition of deep paths, thereby capturing the dynamic characteristics of nodes at different frequencies and levels.

First, weighted sum convolution is used to propagate in multiple steps on the network, gradually obtaining node features at each scale; then the neighbor information is aggregated. Let operator $G(\cdot)$ represent a weighted neighbor information aggregation. For the node feature matrix $X \in {R}^{N \times d}$, the propagation can be written as:
\begin{equation}
X^{(t)} = G(X^{(t-1)}), \quad t = 1, 2, \cdots, T
\end{equation}

Where $X^{(0)} = X$ is the initial node feature. Key scale selection: $s_1, s_2, \cdots, s_J$. At these scales we get node information $X^{(s_1)}, X^{(s_2)}, \cdots, X^{(s_J)}$:
\begin{equation}
s_J = 2^{J-1}
\end{equation}

Perform multi-scale stitching on the obtained node information:
\begin{equation}
U = \text{concat}(X^{(s_1)}, X^{(s_2)}, \cdots, X^{(s_J)})
\end{equation}

After extracting node information at each key scale (experimentally set $J$ to 3, i.e., the number of steps is 1, 2, and 4), the difference between consecutive scales is calculated. This step can capture the first-order and second-order feature changes. Additional propagation is performed on $\mathbf{U}$ to obtain the second-order features:

\begin{equation}
\mathbf{U}^{(0)} = \mathbf{U}, \quad \mathbf{U}^{(t)} = \mathcal{G}(\mathbf{U}^{(t-1)}), \quad t = 1, \cdots, M
\end{equation}

where $M$ is the number of retransmissions.
First and second-order differences:

\begin{equation}
\mathbf{F}_1 = \text{concat}(\| \mathbf{X}^{(s_1)} - \mathbf{X}^{(s_2)} \|, \| \mathbf{X}^{(s_2)} - \mathbf{X}^{(s_3)} \|, \cdots, \| \mathbf{X}^{(s_{L-1})} - \mathbf{X}^{(s_L)} \| )
\end{equation}

\begin{equation}
\mathbf{F}_2 = \text{concat}(\| \mathbf{U}^{(s_1)} - \mathbf{U}^{(s_2)} \|, \| \mathbf{U}^{(s_2)} - \mathbf{U}^{(s_3)} \|, \cdots, \| \mathbf{U}^{(s_{L-1})} - \mathbf{U}^{(s_L)} \| )
\end{equation}

Final mapping: These differential features $(\mathbf{X}^{(s_L)}, \mathbf{F}_1, \mathbf{F}_2)$ of different scales are concatenated and fused, mapped to the output space through full connection and activation function, and finally form a multi-scale node representation, denoted as Z.
Combining all the above steps, the overall process of GWT module output can be summarized as follows:
\begin{equation}
Z = \text{PReLU}\left(\text{concat}\left(X^{(s_L)}, 
\underset{F_1}{\underbrace{\text{concat}_{i=1}^{L-1} \left| X^{(s_i)} - X^{(s_{i+1})} \right|}}, 
\underset{F_2}{\underbrace{\text{concat}_{i=1}^{L-1} \left| U^{(t)} - U^{(t-1)} \right|}} \right) W + b \right)
\end{equation}
Where $X^{(t)}$ represents the result of $t$-times weighted aggregation of the initial feature $X$; 
$s_1, s_2, \cdots, s_L$ is the key scale; 
$U^{(t)}$ represents the result after $t$ steps of propagation again to $U$ (concatenated by key scale features); 
$|\cdot|$ means element-by-element absolute difference operation; 
$F_1$ is the first-order differential feature concatenation, 
$F_2$ is the second-order differential feature concatenation; 
$W$ and $b$ are the weights and biases of the fully connected layer, respectively; 
and $\text{PReLU}(\cdot)$ represents the activation function. 
$Z$ is the final multi-scale node representation output by the MG Encoder module.

The MG Encoder module not only reveals the complementarity of the global structure (low-frequency components), 
but also reflects the local subtle changes (high-frequency components), 
enabling the model to capture the deep associations hidden in cross-type multi-hop paths, 
providing a richer and more detailed feature representation for the prediction of drug-target interactions.

After encoding, we use the semantic attention mechanism to fuse these multi-scale features to generate a unified node representation with richer representation capabilities. The feature representation extracted by each GWT first obtains the overall semantic information from the perspective of the node. Then a learnable attention vector is used to score the projection results of each perspective to obtain the attention weight of each perspective. And the weighted summation is performed according to the features from different perspectives to obtain the final fused node representation.
The following is the formula of the semantic attention mechanism:
\begin{equation}
\alpha_i = \frac{\exp\left(a^{T} \left( \frac{1}{N} \sum_{j=1}^{N} \tanh(W z_{i,j} + b) \right) \right)}{\sum_{k=1}^{P} \exp\left(a^{T} \left( \frac{1}{N} \sum_{j=1}^{N} \tanh(W z_{k,j} + b) \right) \right)}
\end{equation}

\begin{equation}
Z_{\text{fused}} = \sum_{i=1}^{P} \alpha_i Z_i
\end{equation}
Where $Z_i \in {R}^{N \times d}$ represents the feature representation of all nodes under the $i$perspective,The $j$-th row $Z_{ij}$ represents the features of the $j$-th node under the $i$-th view. Each feature representation of a view is first transformed linearly and activated, and then averaged in the node dimension to obtain the overall semantic information. The attention score is then calculated using the learnable attention vector $a$ and normalized by \textit{softmax}. Finally, the representations of each view are weighted and fused according to the obtained weight $\alpha_i$ to obtain the fused representation $Z_{\text{fused}}$.

\subsection{Multi-level Contrastive Learning}
% Describe HGCN and how neighbors are aggregated
Multi-level contrastive learning belongs to the category of self-supervised learning. Its main idea is to extract the intrinsic structural characteristics of the data by bringing the representations of the same sample under different perspectives or different enhancement methods closer, while pushing the representations of different samples further apart. This method effectively alleviates the dependence on large-scale labeled data and provides high-quality self-supervisory signals for complex network tasks. In the drug-target prediction task, the model extracts features using local neighbor information (SC Encoder module) and deep associations in cross-type and multi-hop paths (MG Encoder module), and then aligns node representations from different perspectives through multi-level contrastive learning, strengthening the complementarity of the two types of information, thereby improving the robustness and generalization ability of the model.

In the contrastive learning process, a key step is to select positive sample pairs. Use multiple deep paths (Such as drug-drug, drug-protein-drug, etc.) to calculate the corresponding similarity matrices, normalize each matrix, and then add them to the unit matrix (ensuring that the similarity of each node with itself is the maximum value), so as to obtain a final similarity matrix that integrates multiple association information.For each row in the similarity matrix (representing a node), if the number of candidate nodes exceeds the preset threshold (for example, the 10 most similar positive sample pairs are selected for drug nodes, and 5, 20, and 20 positive sample pairs are selected for protein, disease, and side effect nodes, respectively), the similarity values of the candidate nodes are sorted in descending order, and the highest number of similarities are selected as positive sample pairs; otherwise, all candidate nodes are directly selected as positive sample pairs. The selection results are recorded in a binary matrix pos\_dict as the basis for alignment of positive sample pairs in contrastive learning.

The losses of the two directions of contrastive learning are unified into one formula. For the h-th layer, the nodes are represented as:
\begin{equation}
S_{ij}^{(h)} = \frac{ \mathbf{Z}_{mg\,j}^{(h)} \cdot \mathbf{Z}_{sc\,j}^{(h)} }{ \left\| {Z}_{mg\,j}^{(h)} \right\| \cdot \left\| \mathbf{Z}_{sc\,j}^{(h)} \right\| \, \tau }
\end{equation}
$z^{(n)}_{\text{mg},j}$ and $z^{(n)}_{\text{sc},j}$ are projection representations from different perspectives. 
$\tau$ is the temperature coefficient set to 0.5. 

We use the positive sample dictionary $P$ to indicate which samples are positive samples 
(i.e., $P_{ij}$ indicates that nodes $i$ and $j$ constitute positive samples), and the loss function is:

\begin{equation}
L^{(h), \mu} = -\frac{1}{n} \sum_{k=1}^{n} \log 
\frac{ \sum_{l=1}^{n} \exp\left( s_{kl}^{(h), \mu} \right) P_{kl}^{(\mu)} }
     { \sum_{l=1}^{n} \exp\left( s_{kl}^{(h), \mu} \right) }
\end{equation}
Where $\mu$ indicates the comparison direction. 
For $\mu = \text{mg} \to \text{sc}$, we define
$s_{ij}^{(h), \text{mg} \to \text{sc}} = s_{ij}^{(h)}$
to represent the similarity from view mg to sc.

The entire multi-level contrastive learning can be written as:

\begin{equation}
L_{\text{contrast}} = \sum_{h=1}^{H} \frac{1}{2} 
\left[ L^{(h), \text{mg} \to \text{sc}} + L^{(h), \text{sc} \to \text{mg}} \right]
\end{equation}
\subsection{Drug Target Prediction}
After obtaining the feature representations of various nodes (such as drugs, proteins, diseases, and side effects), we use the DistMult module to jointly model multiple relationships in the entire heterogeneous network to learn more discriminative entity embeddings. The DistMult model can use the diagonal relationship matrix to model multiple interactions between entities and enable entity embeddings to simultaneously reflect multiple relationship information in joint training. The specific process can be divided into the following parts:

Diagonal matrices as relationship modeling tools: The DistMult model defines a learnable diagonal matrix for each edge relationship. These diagonal matrices only retain the elements on the diagonal when parameterized, which makes the model parameter count small. At the same time, each embedding dimension can be scaled independently to capture the importance of each dimension in different relationships. For any two types of node relationships, we reconstruct them, and their interactions can be approximated by bilinear operations. For example, for drug-protein interactions, the model is reconstructed using the following formula:
\begin{equation}
G_{\text{re}}^{\text{dti}} = \Phi_{\text{drug}} \, \Theta_{\text{dti}} \, \Phi_{\text{protein}}^{T}
\end{equation}
Where $\Phi_{\text{drug}}$ and $\Phi_{\text{protein}}$ represent drug and protein embedding representations, respectively. 
$\Theta_{\text{dti}}$ is a learnable diagonal matrix has non-zero values only on the diagonal, this modeling approach is equivalent to weighting each embedding dimension separately and then multiplying the results, which can capture the contribution of different dimensions in modeling drug-protein interactions. For other relationships in the heterogeneous network, a similar reconstruction method is also used.
For each relationship reconstruction, we calculate the mean square error between the original relationship matrix and the reconstructed relationship matrix as the reconstruction loss. Specifically, for drug-protein interaction:
\begin{equation}
L_{\text{dti}} = \left\| G^{dti} - G^{dti}_{re} \right\|^2 \tag{15}
\end{equation}

The other relations are calculated in the same way, and then the reconstruction loss of all relations is added up as the reconstruction loss:

\begin{equation}
L_{re} = \sum_{r \in R} L_r \tag{16}
\end{equation}

Finally, our model not only obtains multi-view fusion node representation through contrastive learning, but also reconstructs various relationships in heterogeneous networks using the DistMult model. The overall final loss consists of three parts:

\begin{equation}
L = L_{re} + \lambda_1 L_{cl} + \lambda_2 L_{L2} \tag{17}
\end{equation}

$L_{re}$ is the reconstruction loss, $L_{cl}$ is the contrast loss, $L_{L2}$ is the regularization term, $\lambda_1$ and $\lambda_2$ are hyperparameters that control the weights of each loss.
\section{Experimental results and discussion}
\subsection{Baseline}
We compare our model with other models, including FRoGS, SiamDTI, and HyperAttentionDTI.

FRoGS transforms gene representation from traditional identity coding to functional embedding that integrates gene ontology and RNA-seq data, uses hypergraph random walks and contrastive learning to obtain high-dimensional functional vectors, and then performs weighted aggregation on multiple gene embeddings to generate a consensus representation. Finally, it uses a twin neural network to match the transcriptional features generated by compound perturbations with gene functional embeddings, and combines pharmacological activity data for multimodal fusion, thereby achieving accurate drug-target interaction prediction\cite{RN8}.

SiamDTI (Siamese Drug–Target Interaction prediction) is a new method for drug-target interaction (DTI) prediction. Its main feature is the use of a cross-field fusion strategy that combines the local and global structural information of proteins to enhance the model's ability to predict new drugs and targets\cite{RN2}.

HyperAttentionDTI builds a sequence-based end-to-end learning framework, uses a stacked one-dimensional convolutional network (1D-CNN) to extract semantic features from drug SMILES expressions and protein sequences, and designs a new multi-dimensional attention mechanism (HyperAttention) to collaboratively model the complex non-covalent interactions between atoms and amino acids in the spatial and channel dimensions, thereby improving the model's ability to represent drug-target interactions. \cite{zhao2022hyperattentiondti}Compared with the traditional attention mechanism, HyperAttention locates key interaction areas more accurately and has stronger model interpretability.

\subsection{Experimental setup parameters}
We specified GPU for training, and the number of training rounds was set to 5000 rounds; the node embedding dimension was 2048; the learning rate was 0.001; an early stopping mechanism of 500 rounds was adopted to prevent overfitting; at the same time, the random inactivation rates of features and attention were changed from 0.5 and 0.2 respectively; the temperature parameter was set to 0.5 to adjust the cosine similarity calculation; the ratio of positive and negative samples in the drug target information dataset predicted in the experiment was set to 1:10;

We built a model based on a heterogeneous network and generated a drug-protein interaction score matrix, in which each element represents the model's predicted score for the interaction probability of the corresponding drug-target pair. The original dataset consists of multiple drug-target pairs, and each record includes a drug index, a target index, and its interaction label. In the data partitioning process, stratified ten-fold cross-validation was used to ensure that the ratio of positive and negative samples in each fold was consistent with the overall dataset. In each iteration, one fold was used as a test set, and the remaining nine folds were combined into a training set. For model tuning and performance evaluation, the training set was further divided into a training subset (90\%) and a validation subset (10\%).

In the experiment, we use AUC and AUPR as evaluation indicators: AUC refers to the area under the ROC curve, which measures the model's ability to distinguish positive and negative samples under all possible thresholds. AUPR refers to the area under the precision-recall curve, which is mainly used to evaluate the performance of the model under the imbalance of positive and negative samples.

\subsection{Experimental Results}
We run the processed data set under various models. We mainly judge the experimental results based on AUC and AUPR. The experimental results are shown in Figure \ref{fig:Figure_1}:
\begin{figure}[H]
    \centering
    \includegraphics[width=0.9\textwidth]{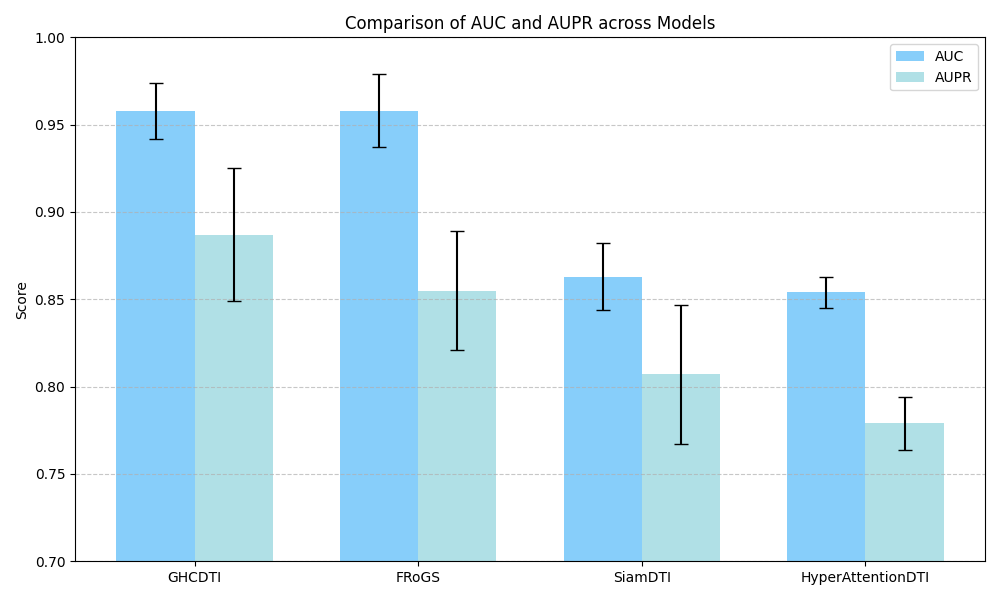}
    \caption{\label{fig:Figure_1}Results of running four models on the dataset}
    \label{fig:Figure_1}
\end{figure}
We ran the models on the same dataset, and GHCDTI showed the best performance. It performed well in correctly distinguishing positive and negative samples and maintaining overall prediction accuracy, and the results were relatively stable. FRoGS followed closely behind. Although its overall performance was slightly inferior to GHCDTI, it still had strong discrimination capabilities. In contrast, SiamDTI and HyperAttentionDTI performed relatively poorly. They had certain deficiencies in identifying real interactions and avoiding misjudged negative samples, and their overall performance was slightly weak.
\begin{figure}[htbp]
  \centering
  \subfigure[]{
    \includegraphics[width=0.45\textwidth]{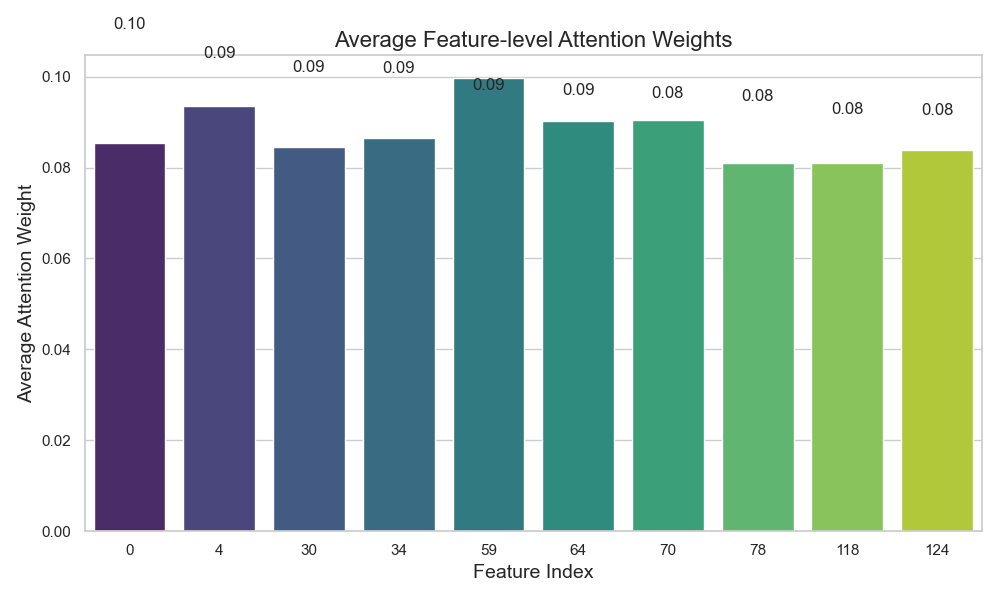}
  }
  \subfigure[]{
    \includegraphics[width=0.45\textwidth]{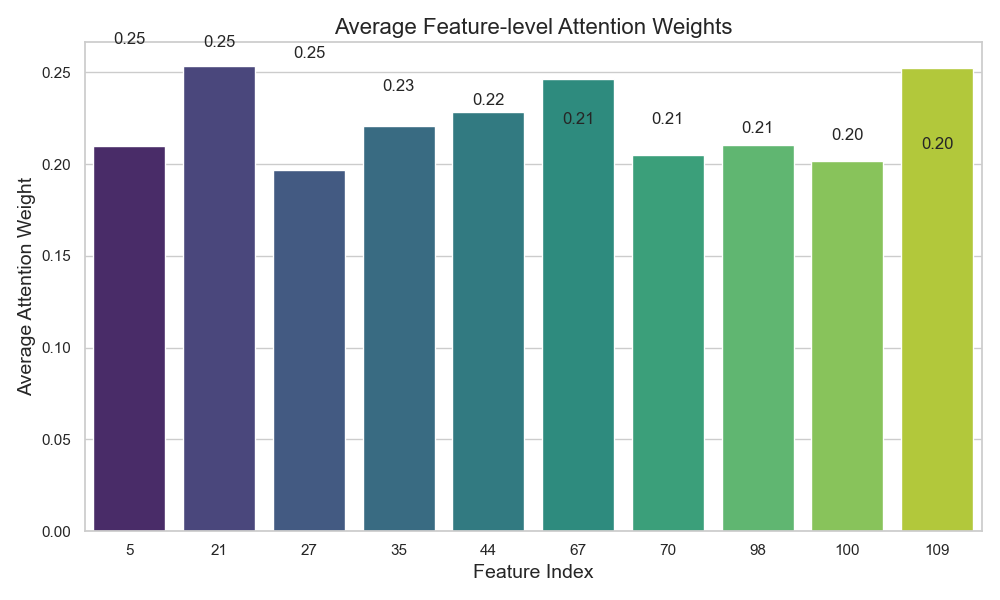}
  }
  \subfigure[]{
    \includegraphics[width=0.45\textwidth]{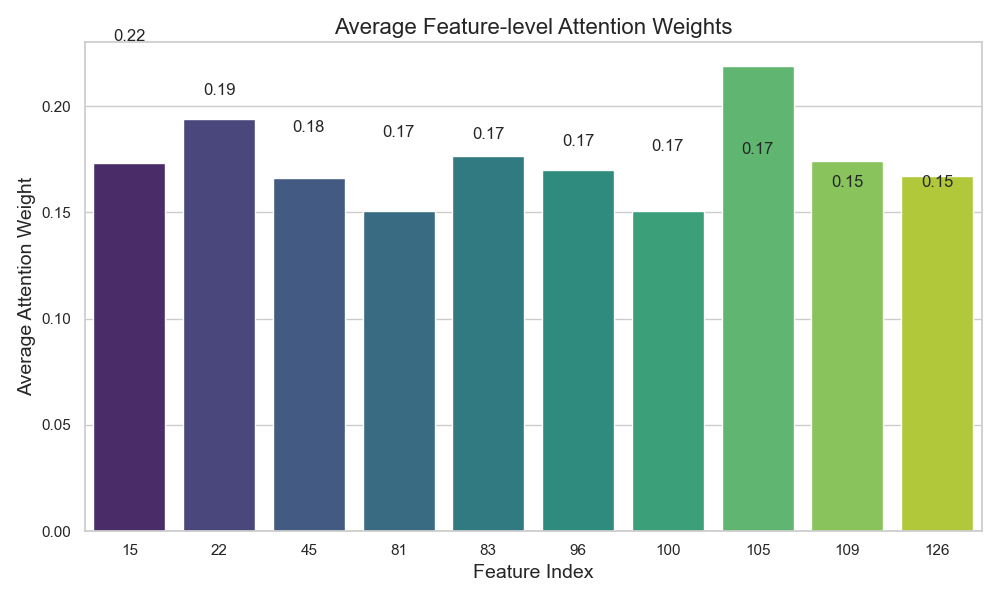}
  }
  \subfigure[]{
    \includegraphics[width=0.45\textwidth]{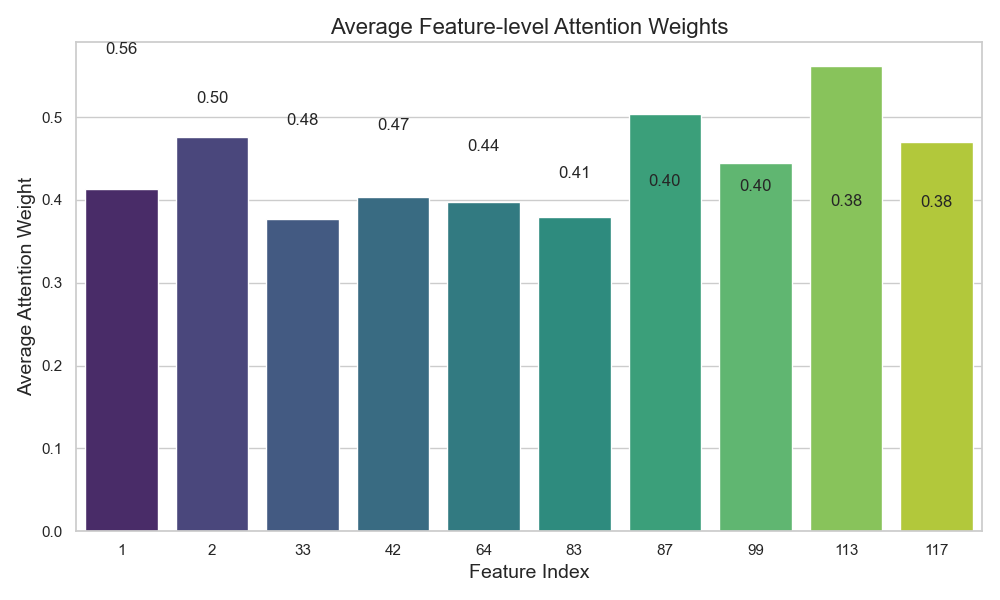}
  }
  \caption{\label{fig:attention-weights}Distribution of attention weights assigned to four node perspectives}
  \label{fig:attention-weights}
\end{figure}

In terms of attention, we have counted the attention weight information from different perspectives, as shown in Figure \ref{fig:attention-weights}. It can be seen that the four perspectives of drugs, proteins, diseases and side effects each provide different information dimensions. Through the attention mechanism, the model can automatically capture the most useful features from each perspective and assign different importance to them.
\subsection{Ablation Experiment}
To verify the contribution of each module in the framework and the sensitivity of key parameters, we systematically designed the following ablation experiments:
Module removal experiment: Control group: complete model (HGCN+GWT+contrastive learning); Experimental group 1: remove the GWT module, only retain HGCN and contrastive learning; Experimental group 2: remove the contrastive learning module, retain HGCN and GWT; Key parameter sensitivity analysis: Experimental group 3: single-layer HGCN; Experimental group 4: GWT scale selection J=2;
The results obtained from the experiment on the dataset are shown in Figure \ref{fig:Figure_2}:

\begin{figure}[H]
    \centering
    \includegraphics[width=\textwidth]{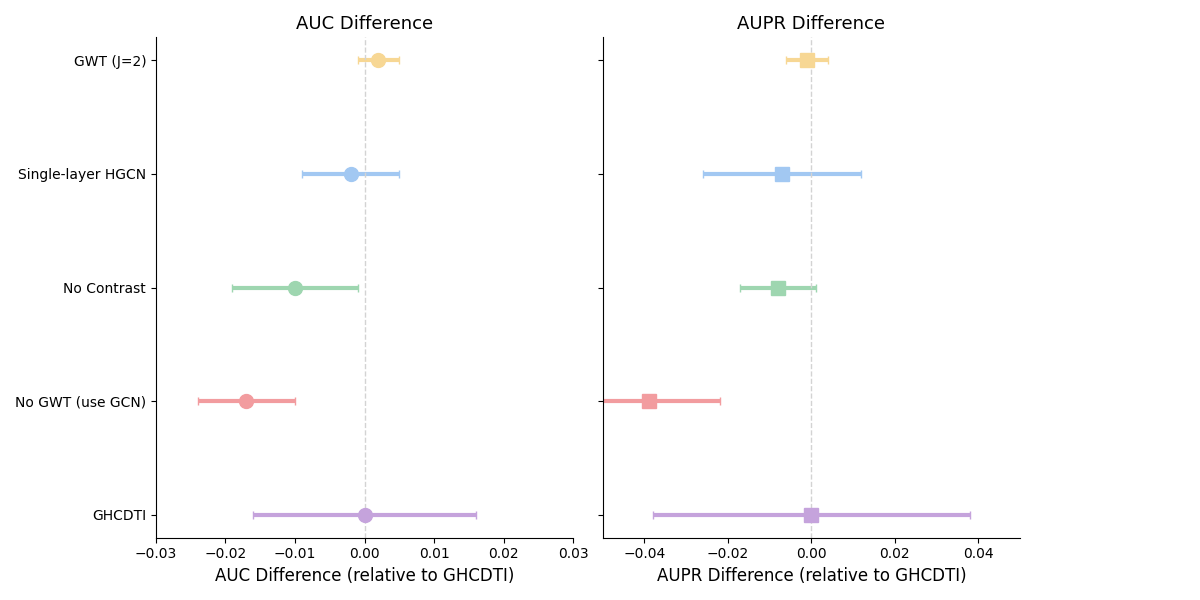}
    \caption{\label{fig:Figure_2}Ablation experiment results after removing parts of the model}
    \label{fig:Figure_2}
\end{figure}

After removing the GWT module, the AUC decreased by about 1.77\% and the AUPR decreased by about 4.40\%; the introduction of the GWT module significantly improved the model's ability to capture complex interactions, and its multi-scale decomposition can better reflect the local changes and global structural information between drugs and targets; after removing the contrastive learning module, the AUC decreased by about 1.04\% and the AUPR decreased by about 0.90\%; the multi-level contrastive learning mechanism helps to align node representations from different perspectives (such as neighbor information and multi-scale features), thereby enhancing the model's ability to distinguish.

When the GWT multi-scale key parameter is changed to 2, the AUPR decreases slightly. Compared with the single-layer structure, the multi-layer HGCN can make better use of multi-hop neighbor information, making the node features more representative. Although the performance of the single-layer HGCN is relatively close, the multi-layer structure has a slight advantage in improving prediction accuracy overall.

These results show that each key component contributes to the improvement of model performance to a certain extent.

\section{Conclusion}
This study constructed a new heterogeneous network drug-target interaction prediction model, which cleverly combined graph wavelet transform and multi-level contrastive learning technology to achieve multi-scale capture and deep analysis of the complex interaction relationship between drugs and targets. The model not only achieved significant improvements in prediction accuracy by aggregating local neighborhood information and decomposing global structural features, but also provided an intuitive perspective for explaining the intrinsic connections between biological molecules. Experimental results show that compared with traditional methods and existing advanced models, this method exhibits higher robustness and stability in all evaluation indicators. It provides a new path for precision drug development and target discovery that takes into account both efficient prediction and mechanism decoding, which has important theoretical significance and practical application prospects.
% Summary of contributions
% Potential impact on interpretable drug discovery
% \clearpage
%\bibliographystyle{splncs04}
% \bibliography{dwf}

\footnote{This work was supported by the Science and Technology Project of Jiangxi Provincial Department of Education, GJJ2201043.}
\end{document}